\documentclass[aps,prl,twocolumn,noshowpacs]{revtex4-1}
\usepackage[utf8]{inputenc}
\usepackage{amsmath}
\usepackage{amssymb}
\usepackage{graphicx}
\usepackage{xcolor}

 \usepackage{units}
\usepackage{hyperref}

\begin{document}

\title{Staircase in magnetization and entanglement entropy of spin squeezed condensates}
\author{H. M. Bharath, M. S. Chapman and C. A. R. S\'a de Melo}
\date{\today}
\affiliation{School of Physics, Georgia Institute of Technology}

\begin{abstract}  
Staircases in response functions are associated with physically observable quantities that respond discretely to continuous tuning of a control parameter. A well-known example is the quantization of the Hall conductivity in two dimensional electron gases at high magnetic fields. Here, we show that such a staircase response also appears in the magnetization of spin-1 atomic ensembles evolving under several spin-squeezing Hamiltonians. We discuss three examples, two mesoscopic and one macroscopic, where the system’s magnetization vector responds discretely to continuous tuning of the applied magnetic field or the atom density, thus producing a magnetization staircase. The examples that we consider are directly related to Hamiltonians that have been implemented experimentally in the context of spin and spin-nematic squeezing. Thus, our results can be readily put to experimental test in spin-1 ferromagnetic $^{87}$Rb and anti-ferromagnetic $^{23}$Na condensates. 
\end{abstract}
\maketitle

In the integer quantum Hall effect, the Hall conductivity changes discretely to continuous tuning of the magnetic field \cite{PhysRevLett.49.405, 0022-3719-15-22-005}. In general, when a system responds discretely to a continuous change of a control parameter, a staircase structure appears in its response function, which is a distinctive characteristic of quantization.  Such phenomenon is significant on two counts. First, one can stabilize the system on a step of the staircase, that is, the flat region between two discrete jumps. Second, these stable states are potentially topological and may carry topological invariants of the system's phase space. The quantum Hall effect has been observed in fermionic two-dimensional (2D) electron gases \cite{RevModPhys.58.519, RevModPhys.71.S298}.

Bosonic analogues of quantum Hall states have been predicted to exist in rotating, weakly interacting Bose-Einstein condensates (BEC)\cite{PhysRevLett.84.6, PhysRevLett.87.120405, PhysRevLett.89.120401,  Cooper_review, 0953-8984-20-12-123202, PhysRevLett.111.090401, PhysRevA.89.013623}. A spinless, non-inteacting, rotating BEC in a harmonic trap is characterized by Landau levels, similar to a 2D electron gas in a magnetic field \cite{PhysRevLett.84.6}. For a rotating BEC, the trap frequency plays the role of the effective magnetic field and the corresponding lowest Landau level is degenerate in the angular momentum about the axis of rotation. This means that there are multiple angular momentum eigenstates within the lowest Landau level, thus, a weak interaction in the system may select one of these angular momentum eigenstates as the ground state of the system depending on the ratio of the interaction strength and the cyclotron frequency \cite{PhysRevLett.84.6, PhysRevLett.87.120405}. Thus, the system's angular momentum responds discretely to continuous tuning of the effective magnetic field, in analogy with the quantum Hall effect. Recently, such phenomena has been predicted even in a spin-1 BEC \cite{PhysRevLett.89.120401} and a pseudo spin-1/2 BEC\cite{PhysRevLett.111.090401}.

For the bosonic examples discussed above, the interaction plays a pivotal role in the emergent angular momentum staircase as a function of the effective magnetic field.  Two other quantum phenomena that also arise from interactions are squeezing and many body entanglement. Spin squeezed states have been prepared in bosonic systems \cite{PhysRevLett.86.5870, PhysRevLett.100.250406, PhysRevLett.104.073602, 2010Natur.464.1165G, 2012NatPhy8305H, Bohnet1297, PhysRevLett.116.093602, Kasevich_2016} and used to enhance the precision in a measurement, for example, of the applied magnetic field. They are characterized by noise in the transverse spin component that is lower than any classical state and are generally prepared with the help of an interaction term in the Hamiltonian. Two of the most common modes of preparing squeezed states, one-axis twisting and two-axis counter twisting, involve interactions \cite{PhysRevA.47.5138}.

In this letter, we show three examples of spin-squeezing Hamiltonians, realizable in spin-1/2 and spin-1 BECs, that are characterized by a staircase response in the magnetization. First we show this for one-axis twisting Hamiltonian.  Second, we demonstrate that an interacting ferromagnetic spin-1 BEC, where spin-nematic squeezing has been demonstrated \cite{2012NatPhy8305H}, also displays a staircase. Third, we consider an interacting anti-ferromagnetic spin-1 BEC, where a staircase is also obtained in the direction of the magnetization. The first two examples are mesoscopic, while the third is a macroscopic phenomenon. We also propose experiments to observe these effects.   

\begin{figure*}[t]
\includegraphics[scale=0.62]{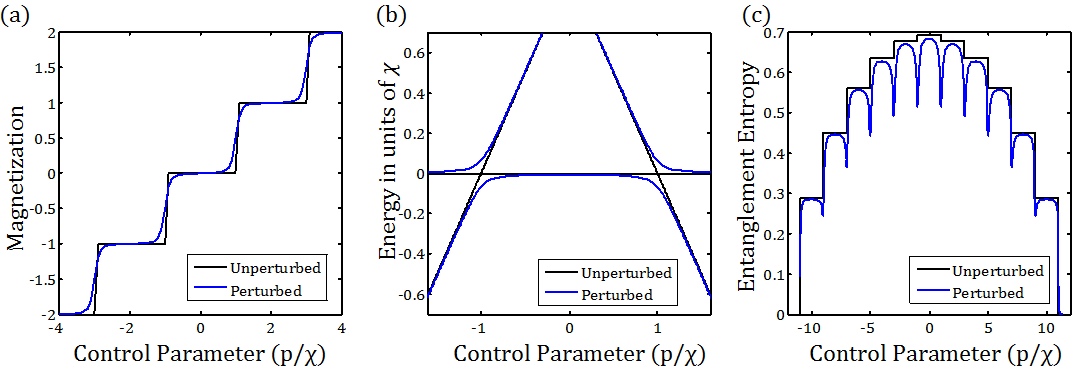}
\caption{\textbf{Staircase in the one-axis twisting Hamiltonian:} (a) Shows the ground state magnetization as a function of the strength of the applied field $p$ for constant interaction $\chi$ in the one-axis twisting Hamiltonian $H= \chi S_z^2 - pS_z$, with (blue curve) and without (black curve) the perturbation $\epsilon S_x$. (b) Shows the corresponding ground state and the first excited state energies around the level crossings between $m=-1$ and $m=0$, as well as $m=0$ and $m=+1$. In the absence of the perturbation, there are true level crossings, but when the perturbation is added, gaps open and thereby smooth the staircase. The term $\epsilon S_x$
is also responsible for changing the system's magnetization, which is otherwise conserved. (c) Shows the entanglement entropy of the local ground state as a function of the control parameter $p/\chi$. The black curve shows the entanglement without the perturbation a written in Eq.\ref{entanglement}, while the blue curve shows the entanglement with the perturbation for 
$\frac{\epsilon}{\chi}=0.02$.}\label{FIG1}
\end{figure*}

\textit{Staircase in one-axis twisting.} First, we consider a pseudo spin-1/2 BEC under the one-axis twisting Hamiltonian, $H= \chi S_z^2$, where $S_z$ is the total spin operator in the $z$-direction and $\chi$ represents the strength of two body interactions in the system \cite{PhysRevA.47.5138}. By applying a magnetic field $p$ in the $z$-direction, we obtain a staircase structure in the ground state magnetization of the Hamiltonian, $H=\chi S_z^2 - p S_z$. We use units where $\hbar=1$, $S_z$ is dimensionless, $\chi$ and $p$ are frequencies. The eigenstates of $S_z$ are also eigenstates of this Hamiltonian. The energy of the eigenstate with a magnetization $m$ is $E_m = \chi m^2 - p m$, for $m= -\frac{N}{2}, -\frac{N}{2}+1, \cdots \frac{N}{2}$, where $N$ is the number of atoms in the condensate. By minimizing the energy, we obtain the ground state magnetization $m_{gs}= [\frac{p}{2\chi}]$, where $[x]$ represents the integer closest to $x$. Here,$\frac{p}{\chi}$ plays the role of the control parameter to which the magnetization responds discretely. The initial step of the magnetization staircase occurs when $\frac{p}{\chi}< -N$, with magnetization $m_{gs}=-\frac{N}{2}$, while the final step occurs when $\frac{p}{\chi}> N$, with magnetization $m_{gs}=\frac{N}{2}$. In between, the $m_{gs}$ responds discretely to continuous variation of $p$ as shown in Fig.~\ref{FIG1}(a).

Every step in this staircase is a distinct quantum state and every jump corresponds to a level crossing. The eigenenergies in the vicinity of a level crossing are shown in Fig.~\ref{FIG1}(b). Notice that this is a true level crossing, even when the system size is small, that is, it is not an avoided level crossing. Consequently, in order to observe this effect, one has to facilitate each jump in the staircase by opening up a gap at the level crossing. This can be done by adding a weak field $\epsilon$ in the $x$-direction leading to the Hamiltonian $H= \chi S_z^2 -p S_z - \epsilon S_x$, where $\epsilon$ is also in units of Hz. The resulting energy gaps for crossings between states with $m_{gs} = -1$ and $m_{gs} =0$, as well as $m_{gs} = 0$ and $m_{gs} = +1 $ are shown in Fig.~\ref{FIG1}(b). The term $\epsilon S_x$ also smoothes out the staircase in Fig.~\ref{FIG1}(a) and is responsible for changing the system's magnetization, which is otherwise conserved.   

The quantum states in this magnetization staircase are related to the familiar Dicke ladder \cite{PhysRev.93.99}, where transitions between neighboring total angular momentum states of atoms can occur coherently leading to superradiance. An experiment where the control parameter $\frac{p}{\chi}$ is slowly swept from $-N$ to $N$ would induce a transfer of the atom population between the spin states, one atom at a time. Furthermore, this is also a way of deterministically producing all the Dicke states in this ladder, most of which are highly entangled \cite{PhysRevLett.115.250502, PhysRevA.88.063802}. In an experiment, the system can be initialized at $m=-\frac{N}{2}$ or $m=\frac{N}{2}$, where it is completely unentangled. As the control parameter $\frac{p}{\chi}$ is tuned, the magnetization $m$ increases in integer steps and the corresponding entanglement entropy also steps up, peaking at $m=0$, see Fig.~\ref{FIG1}(c). The entanglement entropy for magnetization $m$, in terms of the magnetization per atom, $\mu=\frac{m}{N}$, is given by \cite{Supplementary}
\begin{equation}\label{entanglement}
\mathcal{E} = -\left(\frac{1}{2}-\mu\right)\log \left(\frac{1}{2}-\mu\right) - \left(\frac{1}{2}+\mu\right)\log \left(\frac{1}{2}+\mu\right).
\end{equation}
The perturbation $\epsilon S_x$, that was added to maintain adiabaticity at the level crossing, also perturbs the entanglement entropy, as shown in Fig.~\ref{FIG1}(c). The large dips in the entanglement entropy that appear at the level crossings are characteristic of a singular perturbation on the degenerate ground state space. Indeed, at the level crossing between magnetizations $m$ and $m+1$, the unperturbed ground state is a two dimensional space spanned by the eigenstates $\{\vert m \rangle, \vert m+1 \rangle\}$ of $S_z$ with eigenvalues $m$ and $m+1$, respectively. The perturbation breaks this degeneracy and picks one state from this space as the ground state. For instance, with an $\epsilon S_x$ perturbation, the ground state is $\frac{\vert m \rangle - \vert m+1 \rangle}{\sqrt{2}}$, independent of $\epsilon$. This state has a lower entanglement entropy than $\vert m \rangle$ and $ \vert m+1 \rangle$, and it corresponds to the dip in the blue curve in Fig.~\ref{FIG1}(c). Thus, when $\epsilon \rightarrow 0$, the blue curve approaches the black curve at every point, excluding the level crossings.

There are several experimental systems where spin squeezing has been demonstrated using the one-axis twisting Hamiltonian including trapped ion systems \cite{PhysRevLett.86.5870, Bohnet1297}, Bose-Einstein condensates \cite{Strobel424}, double well \cite{PhysRevLett.100.250406} and cavity systems \cite{PhysRevA.66.022314, PhysRevLett.104.073602, 1367-2630-19-9-093021, PhysRevLett.119.213601}. Any of these realizations can be used to observe this effect. In an $^{87}$Rb condensate, this Hamiltonian is realized using the hyperfine levels $\vert F=1, m=0\rangle$ and $\vert F=2, m=-1\rangle$ as the pseudo spin-1/2 states. The squeezing term $S_z^2$ can be produced using a Feshbach resonance \cite{Strobel424} and the linear term, $S_z$ can be generated using microwave dressing. Experiments with $N = 300$ atoms and
 $N\chi \sim 20$ Hz have been demonstrated with a detection noise of $\sim 6$ atoms \cite{Strobel424}. Therefore, detecting this staircase structure is within the realm of current experimental techniques.

The role of the interaction term $\chi S_z^2$ lies in introducing convexity into the energy functional. The energy, $E_m = \chi (m^2 - m \frac{p}{\chi})$ is a convex function in the discrete variable $m$ and the control parameter 
$\frac{p}{\chi}$ contributes a linear term in this function. The minima of a convex function can be shifted by adding a linear term, however these shifts are discontinuous since the variable is discrete. This is the primary characteristic of the ground state energy of Hamiltonian which results in a staircase phenomena. Next, we use this observation to identify a staircase in the magnetization of a ferromagnetic spin-1 BEC, as a second example. 

\textit{Staircase in a ferromagnetic spin-1 BEC.} The Hamiltonian of a ferromagnetic spin-1 BEC of $^{87}$Rb atoms, confined to an optical dipole trap and with an applied magnetic field of $B_z$ along the $z$-direction is \cite{PhysRevLett.81.742} 
\begin{equation}\label{eqn:ferromagnetic-hamiltonian}
\begin{split}
H = &\sum_{i=1}^N \left(-\frac{\hbar^2}{2m}\nabla_i^2 + V_T(\textbf{r}_i)\right) \\
&+ \frac{4\pi \hbar^2}{m}\sum_{i>j} \delta(\textbf{r}_i-\textbf{r}_j)\sum_{F=0,2}\sum_{m_F=-F}^F a_F\vert F,m_F\rangle \langle F,m_F\vert \\
& +\sum_{i=1}^N \left( \mu_B g_F B_z L_{zi} + \frac{\mu_B^2}{\hbar^2\Delta}B_z^2 L_{zi}^2 \right).
\end{split}
\end{equation}
\begin{figure}[b]
\includegraphics[scale=0.68]{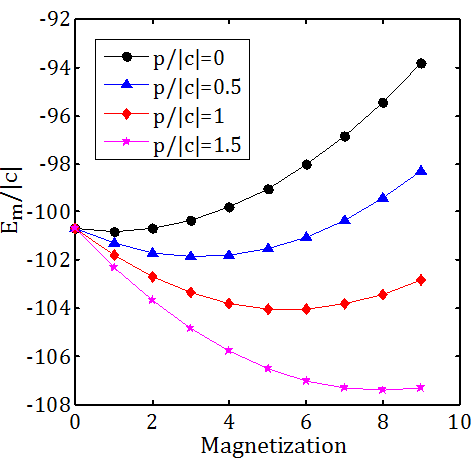}
\caption{\textbf{Convexity of the energy:} The minimum energy eigenvalue 
$E_m$ of the ferromagnetic Hamiltonian $H= c S^2 + qQ_{zz} - pS_z$ is a convex function of the magnetization $m$. For the purpose of this illustration, we have used $N=10$. The minima of these curves correspond to the ground state magnetization. Because $p$ is the coefficient of a linear term in $m$, changing it has the effect of shifting the minimum. The four values of $p/\vert c \vert $ have their minima are different values of $m$, leading to a staircase response of the ground state magnetization as $p/\vert c \vert$ is changed. }\label{FIG2}
\end{figure}
Here, $V_T$ is the dipole trapping potential, the interaction between pairs of atoms is modeled by a $\delta$ function potential and it involves two s-wave scattering lengths, $a_0$ and $a_2$, corresponding to the possible total spin of the two interacting atoms, both of which are in the spin-1 state. In addition, the relevant Land\'e g-factor is $g_F$ and $L_{zi}$ is the spin operator for the $i$-th atom. The hyperfine splitting between the $F=1$ and $F=2$ levels is $\Delta$. Assuming that the trap is sufficiently tight, one can approximate the ground state by a product of a spatial wave function common to all spin modes and a collective $N$-atom spin state. This is also known as the single mode approximation (SMA). Under SMA, the spin part of the Hamiltonian is
\begin{equation}
H= cS^2 + qQ_{zz} - pS_z,
\end{equation}
where $c<0$ is the interaction strength, given by $c= \frac{4\pi\hbar^2 (a_2-a_0)}{3m}\int \vert \phi(\textbf{r})\vert^4 d\textbf{r}$, where $\phi(\textbf{r})$ is the common spatial wave function. The total spin operator of all the atoms is $S^2$,  the strength of the quadratic Zeeman term is $q = \frac{\mu_B^2}{\hbar^2\Delta}$ and the linear Zeeman contribution is $p = \mu_B g_F B_z$ . The collective spin and second rank tensor operators are $S_z = \sum_{i=1}^N L_{zi}$ and $Q_{zz} = \sum_{i=1}^N L_{zi}^2$, respectively. This Hamiltonian has been used to produce spin-nematic squeezed states \cite{2012NatPhy8305H}.

\begin{figure*}[t]
\includegraphics[scale=0.58]{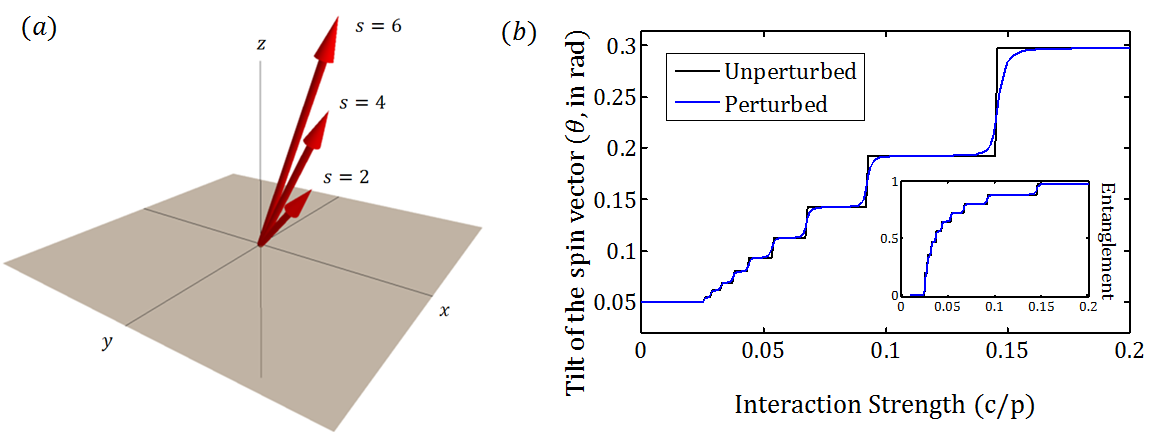}
\caption{\textbf{Staircase in the magnetization direction:} (a) shows the ground state magnetization vector of an anti-ferromagnetic condensate with Hamiltonian $H= cS^2 + pS_x + \alpha Q_{xz}$, for three different values of $c$ with $N=100$. The last term in the Hamiltonian induces the tilting of the magnetization vector by specific angles, depending on where the system is on the staircase. (b) shows the tilt angle for $N=20$ as a function of $c/p$, a staircase, but in contrast with the previous examples, this time it is not only in the magnitude of magnetization, but also in the direction. The blue curve shows the smoothened staircase after adding an $\epsilon Q_{xx}$ perturbation, with $\epsilon = 0.02 p$. The inset shows the ground state entanglement entropy as a function of the control parameter. In both (a) and (b), $\alpha=0.1 p$.}\label{FIG3}
\end{figure*}

We show that the quadratic Zeeman effect induces an energy that is convex in the system's magnetization and therefore, with $c$ and $q$ fixed to appropriate values, we can obtain an analogous staircase in this system. The Hamiltonian commutes with $S_z$ and therefore, it has simultaneous eigenstates with the latter. Let us denote these eigenstates by $\vert n, m\rangle$, with 
\begin{equation}
\begin{split}
(cS^2 + qQ_{zz})\vert n, m\rangle &= \lambda_{nm}\vert n, m\rangle\\
S_z\vert n, m\rangle &= m\vert n, m\rangle\\
\end{split}
\end{equation}
The eigenenergy of this state is $E_{nm}=\lambda_{nm} - p m$. Obtaining the ground state involves a simultaneous minimization over $n$ and $m$. 
We define the function $E_m$ as the minimal value of $E_{nm}$ over all $n$, corresponding to the ground state energy of the Hamiltonian for fixed magnetization. The global ground state is obtained by minimizing $E_m$ over $m$. The Zeeman term $p$ contributes a linear term to $E_m$ leading to
\begin{equation}
E_m = \min_n\{\lambda_{nm}- pm\} = \min_n\{\lambda_{nm}\}- p m
\end{equation}

We use $\vert c \vert$ as our energy unit, and show in Fig.~\ref{FIG2} that $E_m$ is a convex function of $m$. Consequently, the ground state magnetization varies through discrete values of $m$, when the control parameter $p/\vert c \vert$ is tuned. When $q \ll \vert c \vert $, the energy $E_m \approx - \vert c \vert N(N+1) - p m$ is linear in $m$ and has a minimum at $m=\frac{N}{2}$. When $q \gg \vert c \vert $ and $q > p$, the energy $E_m\approx q\vert m \vert - p m $ has a minimum at $m=0$. Upon variation of $q$ between these two extremes, $E_m$ must have a minimum between $m=0$ and $m=\frac{N}{2}$, and must be a convex function of $m$ as seen in Fig.~\ref{FIG2}  .

Thus, we obtain a similar staircase structure in the magnetization, when $\frac{p}{\vert c \vert}$ is varied adiabatically. Like the previous example, the flat areas in the staircase correspond to distinct quantum states and a discrete jump corresponds to a level crossing, which needs to be facilitated by opening up an energy gap. Again, this can be done by perturbing the Hamiltonian with a weak field in the $x$-direction $\epsilon S_x$. In typical experiments \cite{2012NatPhy8305H}, $\vert c\vert \sim 10$ Hz and $q\sim 2\vert c\vert $, indicating that the emergence of the magnetization staircase is also accessible to existing techniques. Similar to the previous example, the entanglement entropy also has a staircase structure.

Both of the examples discussed so far are mesoscopic in the sense that the values of the control parameter corresponding to adjacent steps are separated by $\sim \frac{1}{N}$, where $N$ is the number of atoms. Therefore, in the limit of large atom numbers, it is increasingly more difficult to resolve the different jumps. However, next we show that in an anti-ferromagnetic condensate, a similar staircase structure appears as a truly macroscopic  manifestation, where, the jumps are macroscopically separated.

\textit{Staircase in an anti-ferromagnetic spin-1 BEC.} We consider a spin-1 anti-ferromagnetic BEC with an applied field $p$ in the $z$-direction leading to the Hamiltonian $H = cS^2 - pS_z$, where $c>0$ \cite{PhysRevA.74.033612, PhysRevB.70.115110}. For sufficiently small magnetic field, we can omit the quadratic Zeeman terms. The eigenstates of this Hamiltonian are the total spin states $\vert s, m\rangle$ with $-s \leq m \leq s$ and $s= 0, 2, 4,\cdots, N$ (assuming $N$ is even), due to bosonic symmetry. Here, $s$ is the total spin of the system, that is, $S^2\vert s, m\rangle = s(s+1)\vert s, m\rangle$. The eigenenergy of $\vert s,m\rangle$ is $E_{sm}= cs(s+1)- p m$. When $p>0$, the ground state has $m= + s$. In this case the energy
\begin{equation}
E_{s} = \min_m E_{sm} = cs^2 +(c-p)s
\end{equation}
is a convex function in $s$. In contrast to previous examples, the control parameter is the coefficient $c$ of the quadratic term instead of the field $p$ in the linear Zeeman contribution. 

The ground state value of $s$ is the non-negative integer closest to $\frac{p-c}{2c}$. When $c=0$, the ground state has $s=N$ and when $c\geq p$, it has $s=0$. Because $s$ has a staircase structure, so does the systems magnetization. The level crossings in this staircase occur at values of $c$ given by
\begin{equation}
c_s = \frac{p}{2s-1}; \ \ s= 2, 4, \cdots , N.
\end{equation}
 
The magnetization of the ground state is given by $\langle \vec{S}\rangle = (0, 0, s)$ and develops a staircase structure when $c$ is tuned. We show now that by adding a suitable perturbation to the Hamiltonian, this staircase structure can be transferred to the \textit{direction} of the magnetization. 

Let us perturb the Hamiltonian by $Q_{xz}$, which is a quadratic variable given by $Q_{xz}= \sum_{i}\{L_{xi}, L_{zi}\}$ for a single atom. The Hamiltonian becomes $H= cS^2 - pS_z + \alpha Q_{xz}$. Within a given step in the staircase, 
$\frac{p}{2s+1}< c < \frac{p}{2s-1}$, we use first order perturbation theory to obtain the ground state
\begin{equation}
\vert \psi_s \rangle = \vert s, s\rangle + \frac{\alpha}{p} q_s \vert s, s-1\rangle
\end{equation} 
from the unperturbed ground state $\vert s,s\rangle$.
Here, $q_s = \langle s,s\vert  Q_{xz}\vert s, s-1\rangle = \frac{\sqrt{2s}}{4}\left(\frac{2N+3}{2s+3}\right)$ is the relevant matrix 
element \cite{Supplementary}. In this case, the magnetization 
\begin{equation}
\langle \vec{S}\rangle = s \hat{z} + \frac{\alpha}{p}\sqrt{2s} q_s \hat{x}
\end{equation}
is tilted away from the $z$-axis with 
a polar angle given by 
\begin{equation}
\theta_s = \arctan\left(\frac{\alpha \sqrt{2s}q_s}{p s}\right).
\end{equation}
This angle has a staircase structure with $c$ as the control parameter as shown in Fig.~\ref{FIG3}. Similar to the previous examples, the flat regions of the staircase are distinct quantum states and the associated level crossings need to be facilitated by the opening of a gap created by a perturbation of the type $\epsilon Q_{xx}$, (here, $Q_{xx}=\sum_{i=1}^N L_{xi}^2$) that introduces an overlap between states $\vert s, s \rangle$ and $\vert s \pm 2, s \pm 2 \rangle$. Good candidates to observe this effect experimentally are $^{23}$Na condensates. Typically, $c\sim 20$Hz \cite{PhysRevLett.99.070403} with a macroscopic number of $N=10^5$ atoms. The steps in Fig.~\ref{FIG3}, corresponding to $s=1, 2, 3$, are separated by a few hertz on the $c$ axis and they are independent of the number of atoms. Therefore, this effect is macroscopic and also observable within the existing experimental systems.  



To conclude, we have described three examples where atomic ensembles described by different spin-squeezing Hamiltonians display a staircase structure in their magnetizations as a response to the external tuning of a continuous control parameter. This phenomena can be observed in spin-1 ferromagnetic $^{87}$Rb and anti-ferromagnetic $^{23}$Na condensates, using existing experimental techniques.

\begin{acknowledgments}
\textit{Acknowledgments}: We thank Matthew Boguslawski, Maryrose Barrios and Lin Xin for fruitful discussions. HMB and MSC would like to acknowledge support from the National Science Foundation, grant no. NSF PHYS-1506294. C. A. R. SdM acknowledges the support of the Galileo Galilei Institute for Theoretical Physics via a Simons Fellowship, and the Aspen Center for Physics via NSF Grant No. PHY1607611. 
\end{acknowledgments}

\begin{figure*}[ht!]
\large \textbf{Supplementary Information}
\end{figure*}
\bibliography{Hall_States_Paper}
\bibliographystyle{apsrev4-1}

\newpage

In this supplementary material, we present details of the derivation of 
a few important expressions used in the main text. In section I, we show how the entanglement entropy of many-body systems can be evaluated, and we illustrate our method by using the specific Hamiltonians descirbed in the main text. In section II, we derive the expression for matrix element $q_s= \langle s, s-1 | Q_{xz}|s,s\rangle$ responsible for the staircase structure in the direction of the spin vector.

\section{Entanglement entropy}
A Bosonic many-body state has a unique single atom reduced density matrix, due to its symmetry. Therefore, a convenient measure of many-body entanglement of such states is the Von Neumann entropy of the single atom reduced density matrix. The following simple observation helps us determine the single atom reduced density matrix $\rho$ corresponding to a many-body pure state $|\psi\rangle$ and evaluate its entanglement entropy: If $\hat{o}$ is a single atom observable operator and $\hat{O}=\sum_{i=1}^N \hat{o}_i$ is the corresponding many-body observable, then 
 \begin{equation}\label{exp_values}
 \text{Tr}(\rho \hat{o}) = \frac{1}{N}\langle \psi | \hat{O}|\psi\rangle.
 \end{equation}
For instance, the spin operator along the $x$-axis for a single, spin-1/2 atom is $L_x= \frac{1}{2}\sigma_x$ in units of $\hbar$, where 
$\sigma_x$ is the Pauli matrix. While the many-body spin operator along the $x$-axis is simply given by $S_x = \frac{1}{2}\sum_{i=1}^N \sigma_{xi}$. The symmetry of the many-body state $|\psi\rangle$ ensures that $\text{Tr}(\rho \sigma_x) = \frac{2}{N}\langle \psi | S_x |\psi \rangle $. The reduced density matrix $\rho$ can be reconstructed using the expectation values of a few different observables. 

We illustrate this idea with an example. Let us consider the ground state of the one-axis twisting Hamiltonian $H = \chi S_z^2 - p S_z$, where $S_z$ is the many-body spin operator along the $z$-axis, as described in the main text. The ground state $|\psi\rangle$ of this Hamiltonian is an eigenstate of $S_z$ with a magnetization given by $m =\left[\frac{p}{2\chi}\right]$, the integer closest to $\frac{p}{2\chi}$. It follows from Eq.~\ref{exp_values} that the single atom reduced density matrix $\rho$ of $|\psi\rangle$ has spin expectation values
\begin{equation}
\begin{split}
\text{Tr}(\sigma_x\rho) &= \frac{2}{N}\langle \psi | S_x |\psi \rangle = 0 \\
\text{Tr}(\sigma_y\rho) &= \frac{2}{N}\langle \psi | S_y |\psi \rangle = 0 \\
\text{Tr}(\sigma_z\rho) &= \frac{2}{N}\langle \psi | S_z |\psi \rangle = \frac{2m}{N}. \\
\end{split}
\end{equation}
Using the spin expectation values given above, we may reconstruct the density matrix $\rho$ as
\begin{equation}
\rho = \frac{1}{2}\left(\mathbf{1} + \frac{2m}{N}\sigma_z\right) = 
\left(
\begin{array}{cc}
\frac{1}{2}+\frac{m}{N} &  0 \\
0 & \frac{1}{2}-\frac{m}{N} \\
\end{array}
\right).
\end{equation}
The Von Neumann entropy of this state is $\mathcal{E} = -\text{Tr}[\rho \log (\rho)]$. The explicit expression for $\mathcal{E}$ is given in the main text as Eq.~[1] and shown as the black curve in Fig.~1(c). 

Next, we consider the perturbed Hamiltonian $H = \chi S_z^2 - pS_z - \epsilon S_x$. In the absence of the $\epsilon S_x$ perturbation, a level crossing occurs at $\frac{p}{2\chi}=m+1/2$. Without loss of generality, we may assume that $m \leq \frac{p}{2\chi} \leq m+1$. The ground state magnetization is $m$ when $\frac{p}{2\chi} < m+\frac{1}{2}$ and it is $m+1$ when $\frac{p}{2\chi} > m+\frac{1}{2}$. Accordingly, it is convenient to use $\delta = \frac{p}{2\chi} - m - \frac{1}{2} $ as a function of the control parameter $\frac{p}{\chi}$. The range of $\delta$ is $[-1/2, 1/2]$ and the magnetization switches from $m$ to $m+1$ when $\delta$ crosses zero. The ground state, in the presence of the $\epsilon S_x$ perturbation is a superposition of $|m\rangle$ and $|m+1\rangle$, the eigenstates of $S_z$ with eigenvalues $m$ and $m+1$ respectively
\begin{equation}
|\psi\rangle = u |m\rangle  + v |m+1\rangle.
\end{equation} 
The coefficients $u$ and $ v$ depend on $\delta $ and are given by
\begin{equation}
\begin{split}
u &= \sqrt{\frac{1}{2}\left(1-\frac{\delta}{\sqrt{\delta^2 + C_m^2 \frac{\epsilon^2}{4\chi^2}}}\right)}\\
v &= \sqrt{\frac{1}{2}\left(1+\frac{\delta}{\sqrt{\delta^2 + C_m^2 \frac{\epsilon^2}{4\chi^2}}}\right)}.\\
\end{split}
\end{equation}
Here, $C_m$ is the relevant Clebsch-Gordon coefficient, $C_m = \sqrt{(N/2-m)(N/2+m+1)}$. It is now straightforward to compute the spin expectation values
\begin{equation}
\begin{split}
\text{Tr}(\sigma_x\rho) &= \frac{2}{N}\langle \psi | S_x |\psi \rangle = \frac{2uv C_m}{N} \\
\text{Tr}(\sigma_y\rho) &= \frac{2}{N}\langle \psi | S_y |\psi \rangle = 0 \\
\text{Tr}(\sigma_z\rho) &= \frac{2}{N}\langle \psi | S_z |\psi \rangle =  \frac{2}{N}\left(m + v^2 \right), \\
\end{split}
\end{equation}
which can be used to obtain the reduced density matrix
\begin{equation}\label{Reduced_density_matrix}
\rho = 
\left(
\begin{array}{cc}
\frac{1}{2}+\frac{m+v^2}{N} &  \frac{uv C_m}{N} \\
\frac{uv C_m}{N} & \frac{1}{2}-\frac{m + v^2}{N} \\
\end{array}
\right).
\end{equation}
The off diagonal terms are the largest when $\delta=0$ that is, when the control parameter $\frac{p}{\chi} = 2m+1$. Correspondingly, the Von Neumann entropy of the density matrix defined in Eq.~\ref{Reduced_density_matrix}, has a dip at odd values of $\frac{p}{\chi}$, as shown in Fig.~1(c) of the main text. The same technique can be used to evaluate the reduced density matrix of a many-body spin-1 system, that appears in the second and third examples considered in the main text.

%

\section{The matrix element $\mathbf{q_s}$}
Next, we show how the expression for the matrix element $q_s= \langle s, s-1 | Q_{xz}|s,s\rangle$ used in Eq.~[8] of the main text is derived.  The space of symmetric states of $N$ spin-1 atoms has $\frac{(N+1)(N+2)}{2}$ dimensions. A convenient basis for this space is given by the normalized number states, defined as $|N_+, N_0, N_-\rangle = (a_{-1}^{\dagger})^{N_-}(a_{0}^{\dagger})^{N_0}(a_{+1}^{\dagger})^{N_+}|\text{vac}\rangle$, where $a_i^{\dagger}$ is the creation operator for the $i$-th mode, $|\text{vac}\rangle$ is the vacuum state, with no atoms, and $N=N_++N_0+N_-$ is the total number of atoms.

An alternative basis, also of relevance in the present context is given by the coupled spin states $|s, m\rangle$ with $-s \leq m \leq s$ and $s = N, N-2, \cdots, s_{min}$. When $N$ is even, the minimum value of $s$ is $s_{min}=0$ and when $N$ is odd, $s_{min}=1$. These states are the simultaneous eigenstates of the total spin operators $S^2$ and $S_z$ with eigenvalues given by $S^2 |s,m\rangle = s(s+1)|s,m\rangle$ and $S_z |s, m\rangle = m |s, m\rangle$. 

To evaluate the matrix element of interest, $q_s= \langle s, s-1 | Q_{xz}|s,s\rangle$, we need to determine the action of the operator $Q_{xz}$ on the state $|s,s\rangle$ or the state $|s,s-1\rangle$. However, it is easier to  determine the action of this operator on the number states 
\begin{equation}\label{qxz_number_basis}
\begin{split}
Q_{xz}|N_+&, N_0, N_-\rangle = \\
 c_1|N_+&+1, N_0-1, N_- \rangle - c_2 |N_+, N_0+1, N_--1 \rangle  \\
+ c_3|N_+&-1, N_0+1, N_- \rangle - c_4|N_+, N_0-1, N_-+1 \rangle,\\
\end{split}
\end{equation}
where the coefficients are given by $c_1= \sqrt{\frac{(N_++1)N_0}{2}}$, $c_2= \sqrt{\frac{(N_0+1)N_-}{2}}$, $c_3= \sqrt{\frac{N_+(N_0+1)}{2}}$ and $c_4= \sqrt{\frac{N_0(N_-+1)}{2}}$. 
The result shown in Eq.~\ref{qxz_number_basis} follows from the definition of the many-body operator $Q_{xz} = \sum_{i=1}^N \{L_{xi}, L_{zi}\}$ and the matrix form of the single-atom operator
\begin{equation}
\{L_{xi}, L_{zi}\} = \frac{1}{\sqrt{2}}\left(
\begin{array}{ccc}
0 & 1 & 0 \\
1 & 0 & -1 \\
0 & -1 & 0 \\
\end{array}
\right).
\end{equation} 
The matrix element $\langle s, s-1 | Q_{xz}|s,s\rangle$ can be computed by expressing the coupled spin states $|s, m \rangle $ in the number state basis. Noting that $S_z |s, m\rangle  = m |s, m\rangle$ and $S_z |N_+, N_0, N_-\rangle  = (N_+-N_-) |N_+, N_0, N_-\rangle $, the overlap $\langle s, m |N_+, N_0, N_- \rangle $ vanishes unless $N_+-N_-= m$. Therefore, we may write
\begin{equation}\label{spin_basis_to_num_basis}
|s, s\rangle  = \sum_{k=0}^{\frac{N-s}{2}} A_k |k+s, N-2k-s, k \rangle 
\end{equation}
Since $|s, s-1\rangle = \frac{1}{\sqrt{2s}}S_-|s,s\rangle$ where $S_-$ is the lowering operator, it suffices to determine $A_k$ in order to evaluate the matrix element $q_s$. The coefficients $A_k$ can be evaluated using the observation that the raising operator $S_+$ annihilates the state $|s,s\rangle$, that is $S_+|s,s\rangle =0$. Using the relation
\begin{equation}
\begin{split}
S_+&|N_+, N_0, N_-\rangle =\\
 c_1&|N_++1, N_0-1, N_- \rangle 
+  c_2|N_+, N_0+1, N_--1 \rangle \\ 
\end{split}
\end{equation}
in conjunction with $S_+ |s,s\rangle =0$ gives the recursive relation 
\begin{equation}
A_{k+1} = A_k \sqrt{\frac{(k+s+1)(N-2k-s)}{(k+1)(N-2k-s-1)}}.
\end{equation}
Solving for $A_k$ for $k>0$ we obtain
\begin{equation}\label{coeffs}
A_k = A_0\sqrt{\frac{(k+s)!(N-2k-s-1)!!}{k! (N-2k-s)!!}},
\end{equation} 
where $A_0$ is obtained from the normalization condition $\sum_kA_k^2=1$. It is useful to make the connection with hypergeometric functions here by noticing that the squares of the coefficients $A_k$ are generated by a hypergeometric function, 
\begin{equation}\label{gen_fn}
\sum_{k=0}^{\frac{N-s}{2}}\frac{A_k^2}{A_0^2} x^k= {}_{2}F_1\left(-\frac{(N-s)}{2}, s+1;  -\frac{(N-s-1)}{2}, x\right). 
\end{equation}
This follows from the observation that the above recursion relation on the coefficients $A_k$ is also the recursion between consecutive terms of a hypergeometric series:
\begin{equation}
A_{k+1}^2 = A_k^2 \frac{\left(-\frac{(N-s)}{2} + k\right)(s+1+k)}{\left(-\frac{(N-s-1)}{2} + k\right)}\frac{1}{(k+1)}.
\end{equation}
This gives a simple expression for the first coefficient
\begin{equation}\label{a_0}
A_0 = \frac{1}{\sqrt{{}_2F_1\left(-\frac{(N-s)}{2}, s+1 ;  -\frac{(N-s-1)}{2}, 1\right)}}.
\end{equation}
In fact, there is a closed-form expression for this particular evaluation of hypergeometric functions where the first argument is a negative integer,
\begin{equation}\label{hyp_geo_closed}
{}_2F_1(-n, b; c, 1) = \frac{(c-b)(c-b+1)\cdots (c-b+n-1)}{c(c+1)\cdots (c+n-1)}
\end{equation}
Noting that $\frac{(N-s)}{2}$ is always an integer, Eqs.~\ref{coeffs} and \ref{a_0} together give us all the $A_k$ coefficients. 

We are now ready to evaluate the matrix element $q_s$, which can also be written as 
\begin{equation}
q_s = \frac{1}{\sqrt{2s}}\langle s,s | Q_{xz} S_- |s,s\rangle 
\end{equation}
This expression can be viewed as an overlap between the vectors, $|\psi_1\rangle = Q_{xz}|s,s\rangle$ and $|\psi_2\rangle = S_-|s,s\rangle$, scaled by a factor of $\frac{1}{\sqrt{2s}}$. Besides Eq.~\ref{qxz_number_basis}, another expression that is useful to evaluate this overlap is
\begin{equation}\label{s_minus_number_basis}
\begin{split}
S_-&|N_+, N_0, N_-\rangle =\\
 c_3&|N_+-1, N_0+1, N_- \rangle 
+  c_4|N_+, N_0-1, N_-+1 \rangle, \\ 
\end{split}
\end{equation}
which shows the action of the lowering operator on a number state.  Starting from the expansion of the state $|s,s\rangle$ in the number basis given in Eq.~\ref{spin_basis_to_num_basis}, and using the action of $Q_{xz}$ on a number state given in Eq.~\ref{qxz_number_basis} we may expand $|\psi_1\rangle$ in the number basis. Similarly, $|\psi_2\rangle$ can be expanded in the number basis using Eq.~\ref{s_minus_number_basis}. The overlap $\langle \psi_1 | \psi_2\rangle$ can be written in terms of the coefficients $A_k$ as
\begin{equation}
\sqrt{2s}q_s = \langle \psi_1 | \psi_2\rangle = \sum_{k=0}^{\frac{N-s}{2}}s(N-2k-s+\frac{1}{2})A_k^2 .
\end{equation}
The evaluation of this quantity requires a crucial sum $\sum_{k}kA_k^2$,  which is obtained by taking a derivative of Eq.~\ref{gen_fn}: 
\begin{equation}
\begin{split}
&\sum_{k=0}^{\frac{N-s}{2}}k\frac{A_k^2}{A_0^2} = \\
&\left[\frac{d}{dx}\ {}_{2}F_1\left(-\frac{(N-s)}{2}, s+1 ;  -\frac{(N-s-1)}{2}, x\right)\right]_{x=1}.\\
\end{split}
\end{equation}
The derivative of a hypergeometric function can also be written in terms of a hypergeometric function as
\begin{equation}
\begin{split}
&\frac{d}{dx}\ {}_{2}F_1\left(-\frac{(N-s)}{2}, s+1 ;  -\frac{(N-s-1)}{2}, x\right) = \\
&C_{N,s}\ {}_{2}F_1\left(-\frac{(N-s-2)}{2}, s+2 ;  -\frac{(N-s-3)}{2}, x\right),
\end{split}
\end{equation}
which is a standard relation. In this expression the coefficient  $C_{N, s}$ depends on total number of atoms $N$ and the total spin $s$ as
\begin{equation}
C_{N,s} = \frac{(s+1)\left(\frac{N-s}{2}\right)}{\left(\frac{N-s-1}{2}\right)}.
\end{equation}
We now use Eq.~\ref{hyp_geo_closed} to evaluate this expression at $x=1$ and obtain the desired sum of the series  
\begin{equation}
\sum_{k=0}^{\frac{N-s}{2}} k A_k^2 = \frac{(N-s)(s+1)}{2s+3},
\end{equation}
which leads to the final result for the matrix element
\begin{equation}
q_s = \frac{\sqrt{2s}}{4}\left(\frac{2N+3}{2s+3}\right),
\end{equation}
that we used to determine the staircase in the tilt angle of the spin vector shown in Fig.~3(b) of the main text. 
\end{document}